\documentclass[amssymb,prb,twocolumn,showpacs]{revtex4}

\usepackage{epsfig}
\usepackage{dcolumn}
\usepackage{amsmath}
\begin{document}

\title{Terahertz-induced resistance oscillations in high mobility two-dimensional
electron systems.}

\author{Jes\'us I\~narrea$^{1,2}$ }

\address{$^1$Escuela Polit\'ecnica
Superior,Universidad Carlos III,Leganes,Madrid,28911 ,Spain\\
$^2$Unidad Asociada al Instituto de Ciencia de Materiales, CSIC,
Cantoblanco,Madrid,28049,Spain.}


\begin{abstract}
We report on a theoretical work on  magnetotransport under terahertz radiation with
high mobility two-dimensional electron systems. We focus on
the interaction between the obtained radiation-induced magnetoresistance oscillations (RIRO) and the
Shubnikov-de Haas (SdHO) oscillations.
We study two effects experimentally obtained with this radiation. First, the observed
disappearance of the  SdHO oscillations simultaneously with the vanishing resistance
at the zero resistance states region. And secondly the strong modulation of the SdHO oscillations
at sufficient terahertz radiation power. We conclude that both effects
share the same physical origin, the interference between the
average advanced distance by the scattered electron between
irradiated Landau states, (RIRO),  and the available initial density
of states at a certain magnetic field, (SdHO). Thus, from a physical standpoint, what the
terahertz experiments and theoretical simulations reveal is, on the one hand,  the oscillating
nature of the Landau states subjected to radiation and, on the other hand, how they behave
in the presence of scattering.


\end{abstract}
\maketitle
One of the most striking effects discovered in the last decade regarding radiation-matter
coupling is the microwave-induced magnetoresistance ($R_{xx}$) oscillations (MIRO)\cite{mani1,zudov1}.
This effect shows up in  high mobility
\begin{figure}
\centering \epsfxsize=3.2in \epsfysize=4.3in
\epsffile{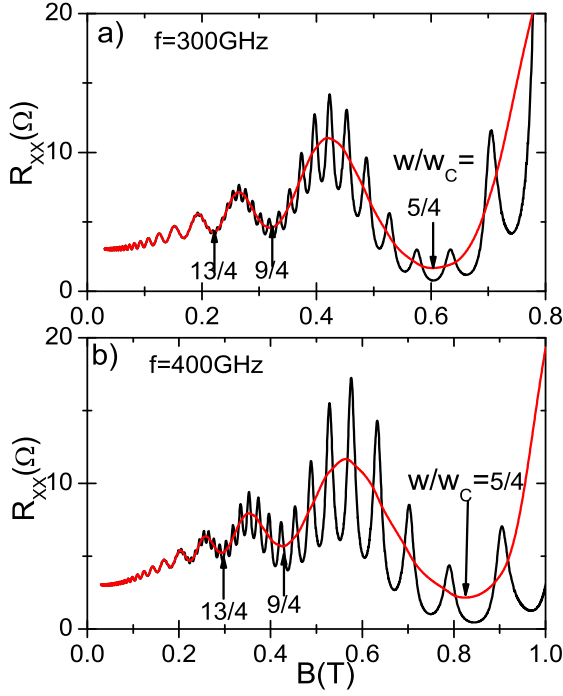}
\caption{ Calculated irradiated  magnetoresistance, $R_{xx}$, vs the magnetic field $B$ for
frequencies of   300 GHz in upper panel and 400 GHz for the lower panel.
Both panels exhibit total  $R_{xx}$,
with radiation-induced resistance oscillations  and Shubnikov-de Haas (SdHO) oscillations, (black curves online), and $R_{xx}$ without
Shubnikov-de Haas (SdHO) oscillations, (red curves online).
}
\end{figure}
two-dimensional electron systems (2DES) when they are irradiated with microwaves (MW) at low
temperature ($T\sim 1K$) and under low magnetic fields ($B$) perpendicular to the 2DES.
Peaks and valleys of MIRO increase with MW  power ($P$) but the latter end up into zero resistance states (ZRS)\cite{mani1,zudov1}
at high enough $P$.
If these effects, MIRO and ZRS, can be qualified as striking, it is even
more remarkable  that after more than ten years of important experimental\cite{mani2,
mani3,willett,mani4,smet,yuan,mani5,wiedmann1,wiedmann2,kons1,vk,mani6,mani61,mani62,mani7,dai}
and theoretical efforts \cite{ina2,ina21,ina22,ina23,ina24,girvin,lei,ryzhii,rivera,vavilov,ina4,ina41,ina42,ina5,ina51}, their physical origin still  remains controversial and far from reaching a definite
consensus among the scientists  devoted to this field.
In this way, the two generally  "accepted"
theoretical models explaining MIRO, (displacement\cite{girvin} and inelastic\cite{vavilov} models)
are completely under question because they are not able to explain either  recent  experimental results\cite{mani71,zudov10} or
the basic features of MIRO.
Among the different features describing MIRO, three of them deserve to be highlighted: first, they are
periodic in $B^{-1}$, second, the oscillations minima present a $1/4$ cycle phase shift and
third, the dependence of MIRO on MW power follows a sublinear relation (square root).
Thus, other theoretical approaches are calling to take over and play a role trying to explain
the different special properties and
 physics of MIRO and ZRS\cite{lei,ina2,ina21}.

In this paper we report on a theoretical work on  magnetotransport in high mobility two-dimensional electron systems
under a higher radiation frequency than
microwaves: terahertz (TH) radiation.
 We focus on
the interaction between the obtained radiation-induced magnetoresistance oscillations (RIRO) and the
Shubnikov-de Haas (SdHO) oscillations. The terahertz band offers the possibility of studying this
interaction because both kind of oscillations coexists while MIRO
tend to vanish when the SdHO are more intense and the effects of interaction are more
difficult to observe.
We first recover the basic or {\it universal} signature of radiation-induced
resistance oscillations:
periodicity in in $B^{-1}$,  the $1/4$ cycle phase shift\cite{mani2,mani4,zudov11} and
 the sublinear radiation power dependence\cite{mani6,mani62}.
Then, we study two new effects experimentally obtained with TH, first the observed
disappearance of the  SdHO oscillations simultaneously with the vanishing resistance
on the zero resistance region\cite{mani90,mani91}. And secondly the strong modulation of the SdHO oscillations
at sufficient TH radiation power\cite{zudov10,ganichev}.
The theoretical contribution of this paper is  based on the {\it radiation-driven electron orbits model}
\cite{ina2,ina21,ina22}.
This model was proposed   to study
the magnetoresistance of a 2DES subjected to MW at low $B$ and temperature, $T$\cite{ina2,ina21,ina30,kerner,park}.
According to this
model, when a Hall bar is illuminated, the
guiding centers of the Landau states perform a classical trajectory consisting in a harmonic
motion along the direction of the current. Thus, the electron orbits move in phase and
harmonically with each other  at the radiation frequency, altering dramatically the scattering
conditions and giving rise eventually to MIRO and, at higher $P$, ZRS.
By using this model, the main result we obtain  is  that these effects are produced by a common origin,  the interplay of the average advanced
scattering distance by the electrons and the initial density of Landau states available at a fixed
 $B$.

\begin{figure}
\centering\epsfxsize=3.2in \epsfysize=5.0in
\epsffile{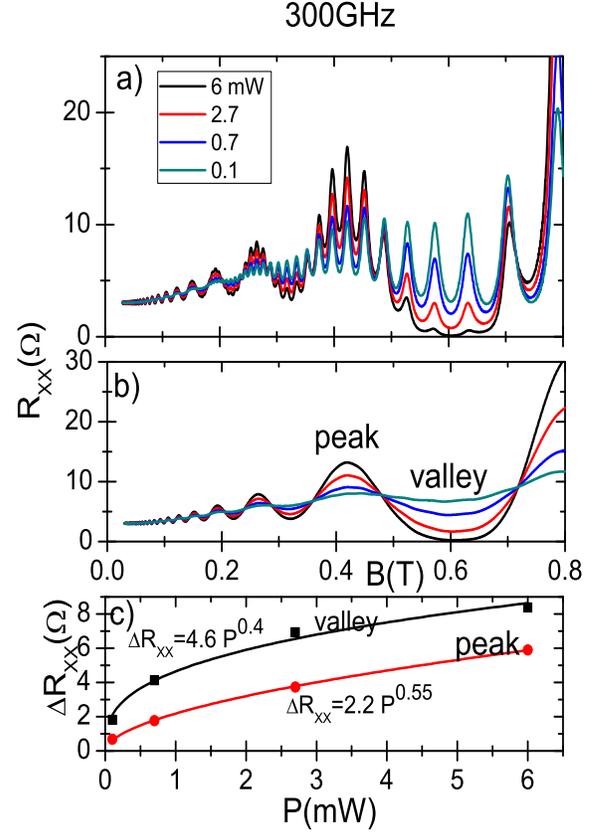}
\caption{Radiation power dependence of the calculated TH-irradiated $R_{xx}$ versus $B$, for increasing
radiation power from to 0.1 mW to 6 mW. Fig. 2a,  total $R_{xx}$ vs $B$,
in 2b,  $R_{xx}$ without Shubnikov-de Haas  oscillations. Fig. 2c exhibits $\Delta R_{xx}$, i.e.,
the difference of irradiated $R_{xx}$ minus the dark one for the labelled peak and valley of 2b,
vs $P$. It is also exhibited the two sublinear fits corresponding to the two sets of values.
}
\end{figure}

As we said above, the  radiation-driven electron orbits model was developed    to analyze
the longitudinal magnetoresistance of a 2DES subjected to radiation\cite{ina2,ina21,ina30,kerner,park}.
Thus, according to this theory, the corresponding time-dependent Schr\"{o}dinger equation can be
exactly solved
and the solution for the total
wave function\cite{ina2,ina30,kerner,park,ina4}reads:
$\Psi_{n}(x,t)\propto\phi_{n}(x-X_{0}-x_{cl}(t),t)$,
where $\phi_{n}$ is the solution for the
Schr\"{o}dinger equation of the unforced quantum harmonic
oscillator. Thus, the  obtained wave
function (Landau state or Landau orbit) is the same as the one of the standard quantum harmonic oscillator where the guiding
center of the Landau state, $X_{0}$ without radiation, is displaced by $x_{cl}(t)$.
$x_{cl}(t)$ is the classical solution of a negatively charged, forced  and damped, harmonic
oscillator\cite{french,main},
\begin{eqnarray}
x_{cl}(t)&=&\frac{-e E_{o}}{m^{*}\sqrt{(w_{c}^{2}-w^{2})^{2}+\gamma^{4}}}\cos ( wt-\beta)\nonumber\\
&=&-A\cos ( wt-\beta)
\end{eqnarray}
 where $E_{0}$ is the amplitude of the radiation electric field and $\beta$ is a phase constant.
$\beta$ is the phase difference between the radiation-driven guiding center and
the driving radiation.
 $\gamma$ is a phenomenologically-introduced damping factor
for the  interaction of electrons with the lattice ions
giving rise to the emission of acoustic phonons.
For high-mobility 2DES\cite{drivenLS}, $\beta\rightarrow \frac{\pi}{2}$, then,
 the time-dependent guiding center is, $X(t)=X_{0}+x_{cl}(t)=X_{0}-A \sin wt$.
This physically implies that the   orbit guiding centers oscillate harmonically at the radiation frequency, $w$.

To calculate the longitudinal conductivity $\sigma_{xx}$ in the 2DES we use the Boltzmann transport theory. With this
theory and within the relaxation time approximation $\sigma_{xx}$ is given by the following equation\cite{ando,miura}:
\begin{equation}
\sigma_{xx}=2e^{2} \int_{0}^{\infty} dE \rho_{i}(E) (\Delta X)^{2}W_{I}\left( -\frac{df(E)}{dE}  \right)
\end{equation}
being $E$ the energy and $\rho_{i}(E)$ the density of
initial Landau states (LS). $W_{I}$ is the remote charged impurity scattering rate, given,  according to
the Fermi's Golden Rule, by
$W_{I}=\frac{2\pi}{\hbar}|<\Psi_{f}|V_{s}|\Psi_{i}>|^{2}\delta(E_{f}-E_{i})$,
where $E_{i}$ and $E_{f}$ are the energies of the initial and final LS.
 $\Psi_{i}$ and $\Psi_{f}$ are the wave functions corresponding to the initial and final LS respectively.
 $V_{s}$ is the scattering potential for charged impurities\cite{ando},
$\Delta X$ is the average distance advanced by the electron between
orbits in every scattering jump in the $x$ direction and is given by,
$\Delta X= \Delta X^{0}-A\sin (2\pi\frac{w}{w_{c}})$.
$\Delta X^{0}$ is the advanced distance without radiation.

\begin{figure}
\centering\epsfxsize=3.2in \epsfysize=4.5in
\epsffile{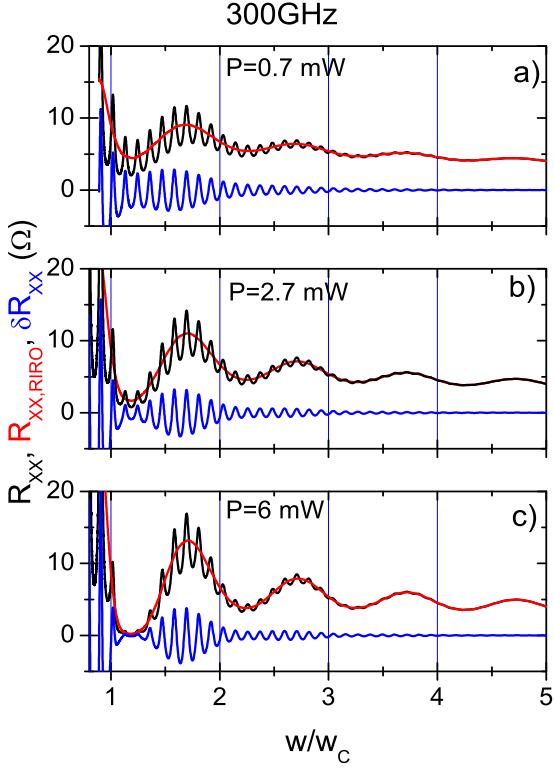}
\caption{Calculated irradiated $R_{xx}$, $R_{xx}$ without SdHO, ($R_{xx,RIRO}$), and the
difference of both, $\delta R_{xx}$   vs $w/w_{c}$ for three different
radiation powers: in panel a) $P=0.7mW$, in panel b) $P=2.7mW$ and in panel c) $P=6.0mW$.
The remarkable result is that SdHO turn out to be modulated by the presence of radiation, being the
modulation harmonic and periodic in $1/B$ and completely in phase with RIRO.
}
\end{figure}
After some algebra we get to an expression for $\sigma_{xx}$\cite{kerner,miura,ina6,ridley}:
\begin{equation}
\sigma_{xx}=\frac{2e^{2} m^{*}}{\pi \hbar^{2}}  (\Delta X)^{2} W_{I} \left[ 1+\frac{2 X_{s}}{\sinh(X_{s})}
   e^{-\frac{\pi\Gamma}{\hbar w_{c}} } \cos\left(\frac{2 \pi E_{F}}{\hbar w_{c}}\right)\right]
\end{equation}
where $X_{s}=\frac{2 \pi^{2}k_{B} T}{\hbar w_{c}}$, $\Gamma$ is the Landau level width
and $E_{F}$ the Fermi energy.
To find the expression for $R_{xx}$ we use
the well-known tensorial  relation
$R_{xx}=\frac{\sigma_{xx}}{\sigma_{xx}^{2}+\sigma_{xy}^{2}}
\simeq\frac{\sigma_{xx}}{\sigma_{xy}^{2}}$, where
$\sigma_{xy}\simeq\frac{n_{e}e}{B}$, $n_{e}$ being  the electron density,  and $\sigma_{xx}\ll\sigma_{xy}$.
Finally, the expression of $R_{xx}$ reads:
\begin{eqnarray}
R_{xx} &\propto& \left[\Delta X^{0}-A\sin \left(2\pi\frac{w}{w_{c}}\right)\right]^{2}\nonumber\\
&\times&  \left[ 1+\frac{2 X_{s}}{\sinh(X_{s})}
   e^{-\frac{\pi\Gamma}{\hbar w_{c}} } \cos\left(\frac{2 \pi E_{F}}{\hbar w_{c}}\right)\right]
\end{eqnarray}
With this expression we want to stand out the terms that are going to be responsible
of the interference between RIRO's, first bracket, and the SdHO, second bracket.

Figure 1 exhibits $R_{xx}$ vs the magnetic field $B$ for a) 300 GHz and b) 400 GHz.
For both panels we represent on the one hand the total $R_{xx}$
(with RIRO and SdHO all together, black curves online). And on the other hand we represent $R_{xx}$ without
SdHO, (red curves online), in order to stand out only the effect of RIRO. By doing this,
we can see intense RIRO in the TH regime that clearly fulfill, for both frequencies two of
the key features of RIRO, such as
the periodicity in $B^{-1}$ and the $1/4$-cycle phase shift of the oscillations minima, ($w/w_{c}=5/4, 9/4, 13/4...$).
Besides,  it is interesting to observe with the TH regime,  how the radiation induced
oscillations overlap with the more rapidly varying with the magnetic field SdHO giving rise to
a strong modulation of the latter. This modulation is explained, according to our model,
by the interference effect between the harmonic terms showing up in Eq. 4. Thus, this
effect is mainly dependent on the radiation frequency and on the Fermi energy or electron density.
In physical terms this interference effect comes  from the interplay of the
average distance advanced by the electron  due
to scattering, that interferes with the available initial density of states.

In Fig. 2, we present the $P$ dependence of the TH irradiated $R_{xx}$ versus $B$ for increasing
$P$, from to 0.1 mW to 6 mW.  In Fig. 2a, we exhibit the complete $R_{xx}$ whereas
in 2b, we plot $R_{xx}$ without SdHO. In Fig. 2c we plot $\Delta R_{xx}$ that is
the difference of irradiated $R_{xx}$ minus the dark one for the labelled peak and valley of 2b,
vs $P$.
For the latter panel, as expected,  we obtain a sublinear dependence of $\Delta R_{xx}$ on $P$.
This can be straightforward explained according to our model since the radiation electric
field $E_{0}$ shows up in the numerator of the amplitude of RIRO and, on the other hand,
$\sqrt{P}\propto E_{0}$. Thus, the exponent of the sublinear expression is close to $0.5$.
But the most interesting effect can be observed in Fig. 2a around $B=0.6$  T.
In this region we obtain the evolution of SdHO as a function of increasing $P$. Interestingly, as
in experiment\cite{mani90}, the SdHO vanish as $R_{xx}$ tends to zero. In other words, we obtain the suppression
of SdHO in the region of radiation-induced zero resistance states. According to our model
this is because this region corresponds to a situation where the advanced distance $\Delta X\rightarrow 0$,
making smaller and smaller the obtained $R_{xx}$, including resistance background and SdHO. Thus,
both simultaneously decrease
in agreement with the experimental results and assessment\cite{mani90}.

In Fig. 3 we present calculated results of irradiated $R_{xx}$, $R_{xx}$ without SdHO, ($R_{xx,RIRO}$), and the
difference of both, $\delta R_{xx}$,   vs $w/w_{c}$ for three different
radiation powers: in panel a) $P=0.7mW$, in panel b) $P=2.7mW$ and in panel c) $P=6.0mW$.
The remarkable result is that SdHO turn out to be modulated by the presence of radiation, being the
modulation harmonic and periodic in $1/B$ and completely in phase with RIRO. It is also noteworthy  that  $\delta R_{xx}$
shows an intense interference effect with the appearance of beats with
increasing intensity for increasing power (see Figs 3a, 3b and 3c). Same results have been previously obtained in experiments\cite{zudov10}.
The coincidence in phase and period is not trivial and reveals deep physical consequences.
Thus, according to our model (Eq. 4), the presence in $R_{xx}$ of the $\Delta X$ term is the main
responsible of the effect. The reason is that $\Delta X$ is harmonically dependent on $w/w_{c}$
 getting this dependence across to the the SdHO term.
In the end, both contributions end up sharing period and phase as obtained in experiments.
In physical terms and as we explain above,  the average advanced  distance of the scattered electron  between radiation-driven LS,
strongly modulates the influence of the initial  density of Landau states on $R_{xx}$.
And this effect can be totally and clearly observed in the TH band and not in the MW
due to the coincidence of  SdHO and RIRO versus $B$, with TH radiation.

In summary, we have reported on a theoretical work on  magnetotransport under terahertz radiation with
high mobility two-dimensional electron systems. We have focused on
the interaction between the obtained radiation-induced magnetoresistance oscillations  and the
Shubnikov-de Haas  oscillations.
We study two effects experimentally obtained with this radiation, first the observed
disappearance of the Shubnikov-de Haas  oscillations simultaneously with the vanishing resistance
on the zero resistance region. And secondly the strong modulation of the Shubnikov-de Haas  oscillations
at high enough TH radiation power. We have applied the radiation-driven electron orbits model
and according to it, both experimental results share the same physical origin:
the interference between the average advanced distance due to scattering between
driven-Landau states, (radiation-induced resistance oscillations), and the available initial density of Landau states,
(Shubnikov-de Haas  oscillations).


This work is supported by the MINECO (Spain) under grant
MAT2014-58241-P  and ITN Grant 234970 (EU).
GRUPO DE MATEMATICAS APLICADAS A LA MATERIA CONDENSADA, (UC3M),
Unidad Asociada al CSIC.


\end{document}